\documentclass[pdf]{UoBnote}
\usepackage[usenames,dvipsnames]{color}     
\usepackage{verbatim}
\usepackage{url}
\usepackage[latin1]{inputenc}
\usepackage[T1]{fontenc}
\usepackage{amsmath, amssymb}
\usepackage{subfigure}
\usepackage{xspace}
\usepackage{multicol}
\usepackage{ftnxtra}
\usepackage{fnpos}

\usepackage{finesse}

\usepackage{fancyvrb}
\DefineVerbatimEnvironment{finesse}{Verbatim}
{formatcom=\small}

\author{Charlotte Bond, Paul Fulda, Andreas Freise}
\shorttitle{Beat coefficients on a bullseye photodiode}
\title{Analytical calculation of Hermite-Gauss and Laguerre-Gauss modes
on a bullseye photodiode}
\date{\today}
\issue{1}
\ligodcc{T1600189}

\usepackage[pdftex,a4paper,pagebackref=true,pdfpagelabels=true]{hyperref}
\definecolor{linkcolor}{rgb}{.8,0,0}
\definecolor{urlcolor}{rgb}{0,0,.7}
\definecolor{citecolor}{rgb}{0,.5,0}
\definecolor{acrocolor}{rgb}{0,0,.7}
\hypersetup{bookmarksopen,colorlinks=true}
\hypersetup{pdfstartview=FitH}
\hypersetup{linktocpage=true,bookmarksnumbered=true}
\hypersetup{plainpages=false,breaklinks=true}
\hypersetup{linkcolor=linkcolor,citecolor=citecolor,urlcolor=urlcolor}

\begin{document}
\maketitle

\tableofcontents 
\vspace{1cm}\hrule \vspace{1.5cm}

\section{Introduction}


Some readers may be familiar with the use of a transversally split photodiode to measure the
misalignment of a beam, effectively detecting the order 1 spatial modes incident on the
photodiode.  Using a similar method a radially split photodiode can be used to measure the mode
mismatch of a beam, by detecting the order 2 spatial modes. 

The idea of using a radially split, or `bullseye' photodiode is not a new 
one; in fact it may be traced back at least to the original `wavefront sensing' paper by Morrison 
et al.~\cite{Morrison1994}. However, it was not until 2000 that Mueller et al. experimentally 
demonstrated the use of bullseye photodetectors in a closed loop for optimizing the mode-matching 
into a Fabry-Perot cavity~\cite{Mueller2000}. 

Recently the issue of mode-mismatch sensing in Advanced LIGO has come to the fore~\cite{brooks2015}. 
Sensing mismatches with bullseye detectors in Advanced LIGO is a significantly more complex challenge than 
in a simple Fabry-Perot cavity, and so accurate simulation of the signals achieved is necessary. For this 
reason it has been necessary to derive the response of such detectors to optical beats between higher-order 
spatial modes. 

This note describes the analytical derivation of the response of bullseye detectors to optical beats between 
higher-order spatial modes of the Laguerre-Gauss form, and subsequently the Hermite-Gauss form. Also 
included is a comparison with numerically calculated beat coefficients, and a simple example of the use of the 
resulting beat coefficients in simulating a mode mismatch sensor for a Fabry-Perot cavity. 

\section{Analytical derivation}

\subsection{The problem}
To correctly calibrate the response of a bullseye photodiode we require the
coefficients for beats between different higher order modes.
We have two fields, the carrier and sidebands, each potentially made
up of a combination of higher order modes.  For the beat between
two higher order modes, $u_{n,m}$ and $u_{n',m'}$, we consider a carrier field:
\begin{equation}
E_{\mathrm{c}}(r,\phi) = a_{n,m} u_{n,m}(r,\phi)
\end{equation}
and a sideband field:
\begin{equation}
E_{\mathrm{sb}}(r,\phi) = a_{n',m'} u_{n',m'}(r,\phi)
\end{equation}
The intensity of a combination of these fields becomes:
\begin{equation}
I(r,\phi) = |E_{\mathrm{c}}+E_{\mathrm{sb}}|^2 = a_{n,m}u_{n,m}a_{n',m'}^*u_{n',m'}^*+ a_{n,m}^*u_{n,m}^*a_{n',m'}u_{n',m'} + |a_{n,m}u_{n,m}|^2 + |a_{n',m'}u_{n',m'}|^2
\label{eqn:Irphi}\end{equation}
We are concerned with the components oscillating at the beat frequency between
the carrier and sidebands. These are the components which include the cross terms,
not the third and fourth terms which are simply DC components.  Thus we can ignore
them in our analysis for a demodulated signal.  The signal generated by a bullseye
photodiode is the difference between the signal generated in the outer and inner areas.
For the terms at the beat frequency this is:
\begin{equation}
\begin{split}
S {}& = a_{n,m} a_{n',m'}^* \int_R^{\infty} \int_0^{2\pi} u_{n,m}u^*_{n',m'} \ r \ \mbox{d}\phi \ \mbox{d}r
	\ + \ a_{n,m}^*a_{n',m'} \int_R^{\infty}\int_0^{2\pi} u^*_{n,m} u_{n',m'} \ r \ \mbox{d}\phi \ \mbox{d}r \\
	{}& - \ a_{n,m} a_{n',m'}^* \int_0^R \int_0^{2\pi} u_{n,m}u^*_{n',m'} \ r \ \mbox{d}\phi \ \mbox{d}r
	\ - \ a_{n,m}^*a_{n',m'} \int_0^R\int_0^{2\pi} u^*_{n,m} u_{n',m'} \ r \ \mbox{d}\phi \ \mbox{d}r \\
{}& = c_{n,m,n',m'} \ a_{n,m} a_{n',m'}^* +  c_{n',m',n,m} a_{n',m'}^*a_{n,m}
\end{split}
\end{equation}
where $c_{n,m,n',m'}$ is the beat coefficient between modes $n,m$ and $n',m'$.

\subsection{Approach}

As the active area of a bullseye photodiode is cylindrically symmetric it is
sensible to approach this analysis using the cylindrically symmetric 
Laguerre-Gauss modes.  There exists a conversion between the Laguerre
and Hermite-Gauss basis, from which we can convert the analytical
result for LG modes to that for the HG modes.

\subsection{Calculating LG beat coefficients}

In terms of LG modes we have the inner integral:
\begin{equation}
k_{p,l,p',l'}^{\mathrm{inner}} = \int_0^R \int_0^{2\pi} u_{p,l} u^*_{p',l'} \ r \mbox{d}\phi \ \mbox{d}r
\end{equation}
and the outer integral:
\begin{equation}
k_{p,l,p',l'}^{\mathrm{outer}} = \int_R^{\infty} \int_0^{2\pi} u_{p,l} u^*_{p',l'} \ r \mbox{d}\phi \ \mbox{d}r
\end{equation}
where $R$ is the radius of the inner circle of the bullseye photodiode.
The Laguerre-Gauss mdoes are given by:
\begin{equation}
\begin{split}
	u_{p,l}(r,\phi,z) {}& = \frac{1}{w(z)} \sqrt{\frac{2p!}{\pi(|l|+p)!}} \exp{(\I(2p+|l|+1)\Psi(z))} \\
	{}& \left(\frac{\sqrt{2}r}{w(z)}\right)^{|l|} L_p^{|l|}\left(\frac{2r^2}{w^2}\right) \exp{\left(-\frac{\I k r^2}{2R_C(z)}-\frac{r^2}{w^2(z)}+\I l\phi\right)}
\end{split}
\end{equation}
where $w$ is the beam spot size, $p$ and $l$ are the radial and azimuthal indices,
$\Psi$ is the Gouy phase, $k$ is the wavenumber and $R_C$ is the radius of curvature of the wavefront.
$L_p^{|l|}$ refer to the associated Laguerre polynomials.

We have:
\begin{equation}
u_{p,l}u^*_{p',l'} = \alpha \left(\frac{\sqrt{2}r}{w}\right)^{|l|+|l'|} L_p^{|l|}\left(\frac{2r^2}{w^2}\right)
	L_{p'}^{|l'|}\left(\frac{2r^2}{w^2}\right) \exp{\left(-\frac{2r^2}{w^2}+\I(l-l')\phi\right)}
\end{equation}
where the complex constant $\alpha$ is given by:
\begin{equation}
\alpha = \frac{2}{w^2\pi} \sqrt{\frac{p!p'!}{(|l|+p)!(|l'|+p')!}} \exp{(\I(2p+|l| - 2p' - |l'|)\Psi)}
\end{equation}
It is convenient to include the Gouy phase explicitly in this form,
because in \textsc{Finesse} simultion the Gouy phase is stored in the
amplitude coefficients for the modes and we need to remove it from the
beat coefficients computation.

The integral can be easily separated into the two variables $r$ and $\phi$.
For the angular integration we simply have:
\begin{equation}
I_{\phi} = \int_0^{2\pi} \exp{(\I(l-l')\phi)} \mathrm{d}\phi = \left[\frac{e^{\I(l-l')\phi}}{\I(l-l')}\right]_0^{2\pi} = 0
\end{equation}
The only non-zero result is achieved when $l=l'$, i.e. when the exponential drops
out before the integration.  In this case:
\begin{equation}
I_{\phi}(l=l') = 2\pi
\end{equation}
This is a useful result as it shows that the beats between any two modes with $l\neq l'$
do not contribute for the case of a bullseye photodiode.  This makes sense as the active
area of the photodiode has no angular dependence.

We now consider the radial integration:
\begin{equation}
I_r = \int_a^b \left(\frac{\sqrt{2}r}{w}\right)^{2|l|} L_p^{|l|}\left(\frac{2r^2}{w^2}\right)
	L_{p'}^{|l|}\left(\frac{2r^2}{w^2}\right)\exp{\left(-\frac{2r^2}{w^2}\right)} \ r \ \mbox{d}r
\end{equation}
where $a$ and $b$ represent the limits for the integration for either the inner
($a=0$, $b=R$) or outer ($a=R$, $b=\infty$) ring and we use $l=l'$ as the only non-zero contribution.
We now make a variable substitution:
\begin{equation}
   \begin{matrix} 
      x = \frac{2r^2}{w^2} & \ \ \ \ \ & \mbox{d}r = \frac{w^2}{4r} \ \mbox{d}x \\
   \end{matrix}
\end{equation}
We now have:
\begin{equation}
I_r = \int_A^B x^{|l|} L_p^{|l|}(x)L_{p'}^{|l|}(x) \exp(-x) \ \mbox{d}x
\end{equation}
where $A$ and $B$ are the new limits of the integral, $A=0$, $B=\frac{2R^2}{w^2}$ for the
inner integral and $A = \frac{2R^2}{w^2}$, $B=\infty$ for the outer integral.
The associated Laguerre polynomials are of the form:
\begin{equation}
L_p^{|l|}(x) = \sum_{i=0}^p (-1)^i \frac{(p+|l|)!}{(p-i)!(|l|+i)!i!}x^i
\end{equation}
We end up trying to integrate a function of the form $x^n e^{-x}$ for each term in
the two Laguerre sums:  
\begin{equation}
I_r = \frac{w^2}{4} (p+|l|)! (p'+|l|)! \sum_{i=0}^p\sum_{j=0}^{p'} (-1)^{i+j} \int_A^B \frac{x^{|l|+i+j}e^{-x}}
		{(p-i)!(p'-j)!(|l|+i)!(|l|+j)!i!j!} \mbox{d}x
\end{equation}
In the case of the limits on the integration for the inner and outer integration,
this integral can be solved using the lower (inner) and upper (outer) incomplete
gamma functions.
The lower incomplete gamma function has the form:
\begin{equation}
\gamma(n,x) = \int_0^x t^{n-1} e^{-t} \ \mbox{d}t = (n-1)!\left(1-e^{-x}\sum_{k=0}^{n-1} \frac{x^k}{k!}\right)
\end{equation}
for integer $n$.
The upper incomplete gamma function is given by:
\begin{equation}
\Gamma(n,x) = \int_x^{\infty} t^{n-1} e^{-t} \ \mbox{d}t = (n-1)! \ e^{-x}\sum_{k=0}^{n-1}\frac{x^k}{k!}
\end{equation}

So the beat coefficients in terms of Laguerre-Gauss modes can be written as:
\begin{equation}
\begin{split}
c_{p,l,p',l'} {}& = k^{\mathrm{outer}}_{p,l,p',l'} - k^{\mathrm{inner}}_{p,l,p',l'} \\
	{}& = \delta_{l,l'}\sqrt{p!p'!(|l|+p)!(|l|+p')!}  \exp{(\I(2p-2p')\Psi)} \left[
		\sum_{i=0}^{p}\sum_{j=0}^{p'} \frac{(-1)^{i+j} \Gamma\left(|l|+i+j+1,\frac{2R^2}{w^2}\right)}
		{(p-i)!(p'-j)!(|l|+i)!(|l|+j)!i!j!} \right. \\
	{}& - \left.
		\sum_{i=0}^{p}\sum_{j=0}^{p'} \frac{(-1)^{i+j} \gamma\left(|l|+i+j+1,\frac{2R^2}{w^2}\right)}
		{(p-i)!(p'-j)!(|l|+i)!(|l|+j)!i!j!} \right]
\end{split}
\end{equation}
where $d_{l,l'}$ is the Kronecker delta, fulfilling the condition of non zero
results only when $l=l'$.

\subsection{Converting to HG beat coefficients}

Any HG mode can be expressed as a sum of LG modes:
\begin{equation}
u_{n,m} = \sum_{p,l} a_{p,l} u_{p,l}
\end{equation}
where the sum over $p$ and $l$ refers to a sum over all the LG modes
with the same order as the desired HG mode, i.e. $n+m=2p+|l|$.
See \cite{phd.bond2014} for a detailed description of this conversion.

Thus we can write the overlap of two HG modes on a bullseye photodiode
in terms of LG modes.  For example, the inner overlap is given as:
\begin{equation}
k^{\mathrm{inner}}_{n,m,n',m'} = \int_0^R \int_0^{2\pi}  u_{n,m}u^*_{n',m'} \ r \ \mbox{d}\phi \ \mbox{d}r
	= \int_0^R \int_0^{2\pi} \sum_{p,l}a_{p,l}u_{p,l} \sum_{p',l'}a^*_{p',l'}u_{p',l'}^*  r \ \mbox{d}\phi \ \mbox{d}r
\end{equation}
We can now use the solutions of the integrals derived in the previous section.
The equation is simplified as the only non-zero contributions come
from the terms in the sum with $l=l'$, as shown in the previous section.  This removes
a lot of terms in the integral and we finally have:
\begin{equation}
k^{\mathrm{inner}}_{n,m,n',m'} = \sum_{l} a_{p,l}a^*_{p',l}k^{\mathrm{inner}}_{p,l,p',l}
\label{eqn:HGbeatcoeffs}\end{equation}
where we sum over only the $l$ values which appear in the expansions of
both Hermite-Gauss modes ($l=l'$).  To determine the range of $l$ we consider
an example beat between two modes of different orders, the HG$_{02}$
mode and HG$_{22}$ mode.  These modes can be expressed in terms
of LG modes or order 2 for HG$_{02}$:
\begin{equation}
\mbox{HG}_{02} = -0.5 \times \mbox{LG}_{0-2} - 0.7071 \times \mbox{LG}_{10}	-0.5 \times \mbox{LG}_{02}
\end{equation}
and order 4 for HG$_{22}$:
\begin{equation}
\mbox{HG}_{22} =	0.25 \times \mbox{LG}_{0-4} + 0.5 \times \mbox{LG}_{1-2} + 0.6124\times \mbox{LG}_{20} 
+ 0.5 \times \mbox{LG}_{12} + 0.25\times \mbox{LG}_{04}
\end{equation}
To calculate the beat between these two modes we need only consider the LG modes
with $l=l'$.  In this example the azimuthal indices common to both expansions
are $l=-2$, $l=0$ and $l=2$.  So we only need these 3 terms in our calculation
of the beat coefficient.  Consider instead the beat between HG$_{02}$ and
the order 3 mode, HG$_{21}$, which can be expressed as:
\begin{equation}
\mbox{HG}_{21} = 0.6124\I \times \mbox{LG}_{0-3} - 0.3536\I \times \mbox{LG}_{1-1} + 0.3536\I \times \mbox{LG}_{11} - 0.6124\I \times \mbox{LG}_{03}
\end{equation}
In this case there are no common $l$ modes between the two modes and
the beat is zero.  This is the case for all modes when the difference between
the orders is odd.  In the case where the difference between the mode
orders is even we have $l=l'=-O_{\mathrm{min}}, -O_{\mathrm{min}}+2, -O_{\mathrm{min}}+4,
\dots O_{\mathrm{min}}$, 
where $O_{\mathrm{min}}$ is the smaller of the two orders.

\section{Example Coefficients}
\subsection{Derived beat coefficients for an optimally sized beam}
The formula given in eqn.~\ref{eqn:HGbeatcoeffs} can be used to calculate the beat coefficients 
as measured by a bullseye photodetector between all HG modes up to a specified order. 
The beat coefficients calculated in this way are printed up to mode order 3 
in table~\ref{tab:coeffstoorder3}, for the specific case where the radius of the inner portion of the 
detector relative to the incident beam radius is chosen such that the HG$_{00}$ $\times$ HG$_{00}$ 
beat coefficient is negligibly small. In practise this should mean that the total power falling inside 
the inner portion of the detector is equal to the total power falling in the outer portion. 

The detector inner portion radius relative to beam radius required to achieve this condition was 
found to be $0.5887050112577$.

\begin{table}
\begin{center}
\begin{tabular}{|c|c|c|c|c|}
\hline
n & m & n$'$ & m$'$ & coeff. \\
\hline   
 0 &  0 &  0 &  2 & 0.490129071734255 \\
 0 &  0 &  2 &  0 & 0.490129071734255 \\
 0 &  0 &  2 &  2 & -0.226460336800429 \\
 0 &  1 &  0 &  1 & 0.693147180560006 \\
 0 &  1 &  0 &  3 & 0.294216182370426 \\
 0 &  1 &  2 &  1 & 0.169865792091509 \\
 0 &  1 &  2 &  3 & -0.159974289570813 \\
 0 &  2 &  0 &  2 & 0.706913350718638 \\
 0 &  2 &  2 &  2 & 0.134069741206516 \\
 0 &  3 &  0 &  3 & 0.711794198742489 \\
 0 &  3 &  2 &  3 & 0.15405930406129 \\
 1 &  0 &  1 &  0 & 0.693147180560006 \\
 1 &  0 &  1 &  2 & 0.169865792091509 \\
 1 &  0 &  3 &  0 & 0.294216182370426 \\
 1 &  0 &  3 &  2 & -0.159974289570813 \\
 1 &  1 &  1 &  1 & 0.933373687519067 \\
 1 &  1 &  1 &  3 & 0.0679783724283818 \\
 1 &  1 &  3 &  1 & 0.0679783724283818 \\
 1 &  1 &  3 &  3 & -0.0688289693357714 \\
 1 &  2 &  1 &  2 & 0.896516597036745 \\
 1 &  2 &  3 &  2 & 0.0967993502022488 \\
 1 &  3 &  1 &  3 & 0.929666955955726 \\
 1 &  3 &  3 &  3 & 0.0707108474910448 \\
 2 &  0 &  2 &  0 & 0.706913350718638 \\
 2 &  0 &  2 &  2 & 0.134069741206516 \\
 2 &  1 &  2 &  1 & 0.896516597036745 \\
 2 &  1 &  2 &  3 & 0.0967993502022488 \\
 2 &  2 &  2 &  2 & 0.920269481591934 \\
 2 &  3 &  2 &  3 & 0.907333325165518 \\
 3 &  0 &  3 &  0 & 0.711794198742489 \\
 3 &  0 &  3 &  2 & 0.15405930406129 \\
 3 &  1 &  3 &  1 & 0.929666955955726 \\
 3 &  1 &  3 &  3 & 0.0707108474910448 \\
 3 &  2 &  3 &  2 & 0.907333325165518 \\
 3 &  3 &  3 &  3 & 0.927829909543617 \\
 \hline
 \end{tabular}
 \caption{Analytically calculated bullseye detector HG beat coefficients up to order 3.}
 \label{tab:coeffstoorder3}
  \end{center}
 \end{table}

\subsection{Comparison with numerically calculated beat coefficients}
In order to test the validity of the analytical results, a numerical calculation of beat coefficients was 
also performed. In this calculation, the transversal amplitude functions of selected HG modes were 
generated to fill a 2D matrix. The coherent sum of two such amplitude functions was calculated and 
`dc terms' were discarded, just as in the steps immediately following eqn.~\ref{eqn:Irphi}. Finally, 
the elements of the resulting matrix were summed for the outer and inner segments of the bullseye detector 
area and the difference between the total contributions in the inner and outer areas found. 

This numerical beat coefficient calculation was performed for a few different combinations of HG modes, 
and for each combination for a range of different grid resolutions. The grid size relative to the beam radius 
was a constant at 10, and was always verified to be large enough such that greater than 99.9\% of the power 
of both modes was within the grid area. 

\begin{figure}[htb]
\includegraphics[width=\textwidth]{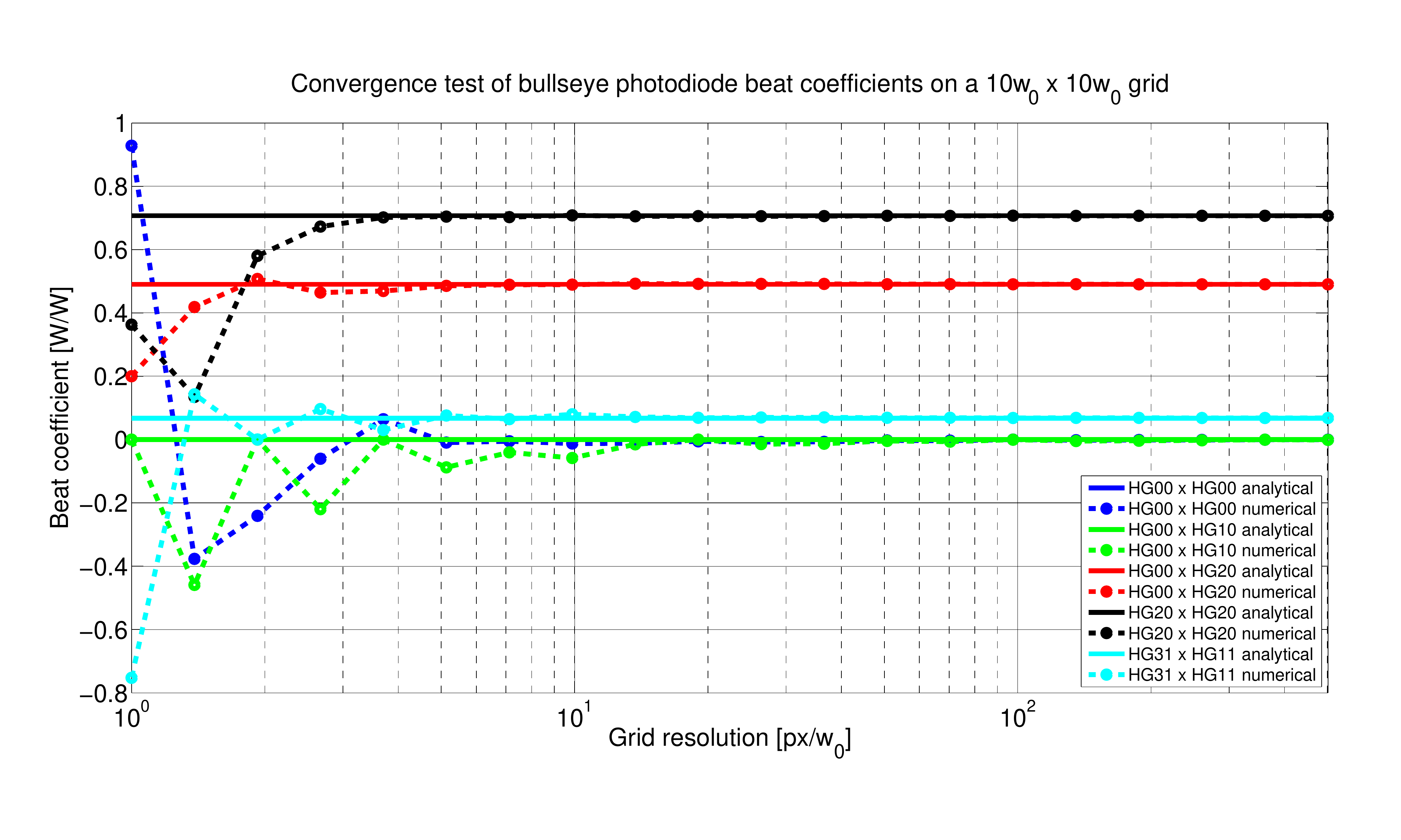}
\caption{A comparison between beat coefficients obtained by the analytical result in eqn.~\ref{eqn:HGbeatcoeffs}, and those obtained by numerical 
integration. As the grid resolution used for the numerical integration is increased, the result tends to that provided by the analytical 
solution.}\label{fig:BPDconverge}\end{figure}

A comparison between the analytically derived results for a few different HG mode beat coefficients 
is shown in Fig.~\ref{fig:BPDconverge}. As the grid resolution increases, the numerical results tend 
towards the analytical results for each of the 5 different HG mode combinations considered. 

\section{Demonstration of bullseye sensing in \textsc{Finesse}}
\begin{figure}[htb]
\begin{center}
\includegraphics[width=0.8\textwidth]{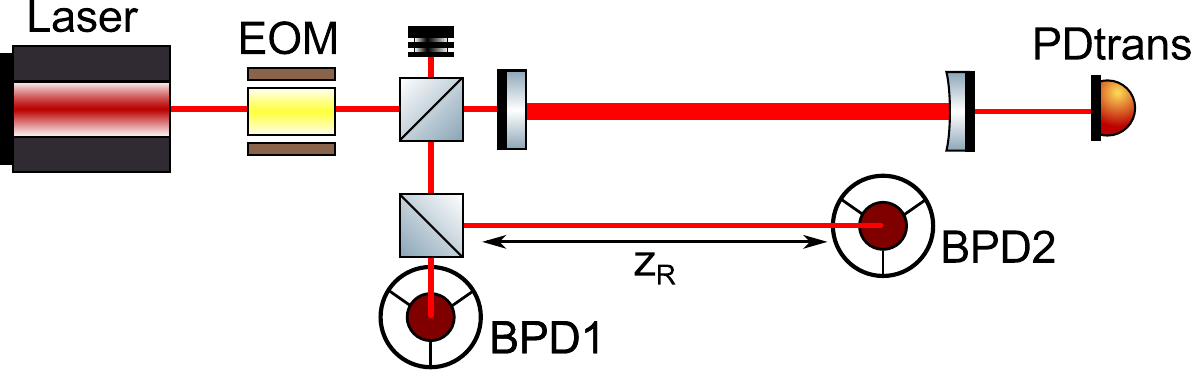}
\caption{Optical layout of the bullseye photodetector simulation example. 
BPD1 is located at the nominal waist location of the cavity eigenmode, 
and BPD2 is one Rayleigh range away from the waist location. 
This gives a 45$^\circ$ Gouy phase separation between BPD1 and BPD2.}
\label{fig:BPDsim}
\end{center}
\end{figure}

In order to see how the derived bullseye detector beat coefficients can be used for the purposes 
of simulating a mode-mismatch sensing scheme, a simple example simulation was performed 
with the software Finesse~\cite{Freise04}. Figure~\ref{fig:BPDsim} shows the optical layout 
of the simulation. 

A laser is phase modulated at 9\,MHz to a modulation index of 0.1 by an EOM, before being made incident 
on a plane-concave Fabry-Perot cavity. The cavity length is 1\,m and the concave end mirror 
has a radius of curvature of 2\,m, giving a cavity eigenmode waist of 582\,$\mu$m located 
at the flat input mirror. The cavity input mirror transmission is 1\% and the end mirror transmission 
is 0.1\%, putting the cavity in the overcoupled regime.

The light reflected from the cavity is picked off by a beam splitter, before being split again by a second beam splitter. 
The transmitted light from the second beam splitter is detected with a bullseye detector BPD1. The 
distance from the cavity input mirror to BPD1 is set to 0\,m, so that BPD1 is effectively located at the 
cavity waist. The reflected light from the second beam splitter is sent to a second bullseye detector, BPD2, 
located at a distance 1\,m from the cavity waist. Since the cavity eigenmode Rayleigh range is 1\,m, this fixes 
the Gouy phase difference between the two bullseye detectors to be 45$^{\circ}$.

The bullseye detectors are demodulated at the 9\,MHz modulation frequency, with a demodulation 
phase that maximizes their sensitivity to any mismatch between the input beam mode and the cavity eigenmode. 
To test the response of the bullseye detectors to a mode mismatch between input beam and cavity mode, the 
beam parameters of the input beam are varied, using the {\ttfamily gauss} command in \Finesse.
The cavity beam parameters remain fixed throughout this variation of the input beam parameter, 
and thus the simulation represents a mode mismatch of the input beam into the cavity. The input 
beam waist location, and waist size (or equivalently Rayleigh range) are varied separately in order 
to observe the different response of BDP1 and BPD2 to these two types of mismatch. The maximum 
HG mode order considered in the simulation was 6.

\begin{center}
\begin{figure}[htb]
\includegraphics[width=\textwidth]{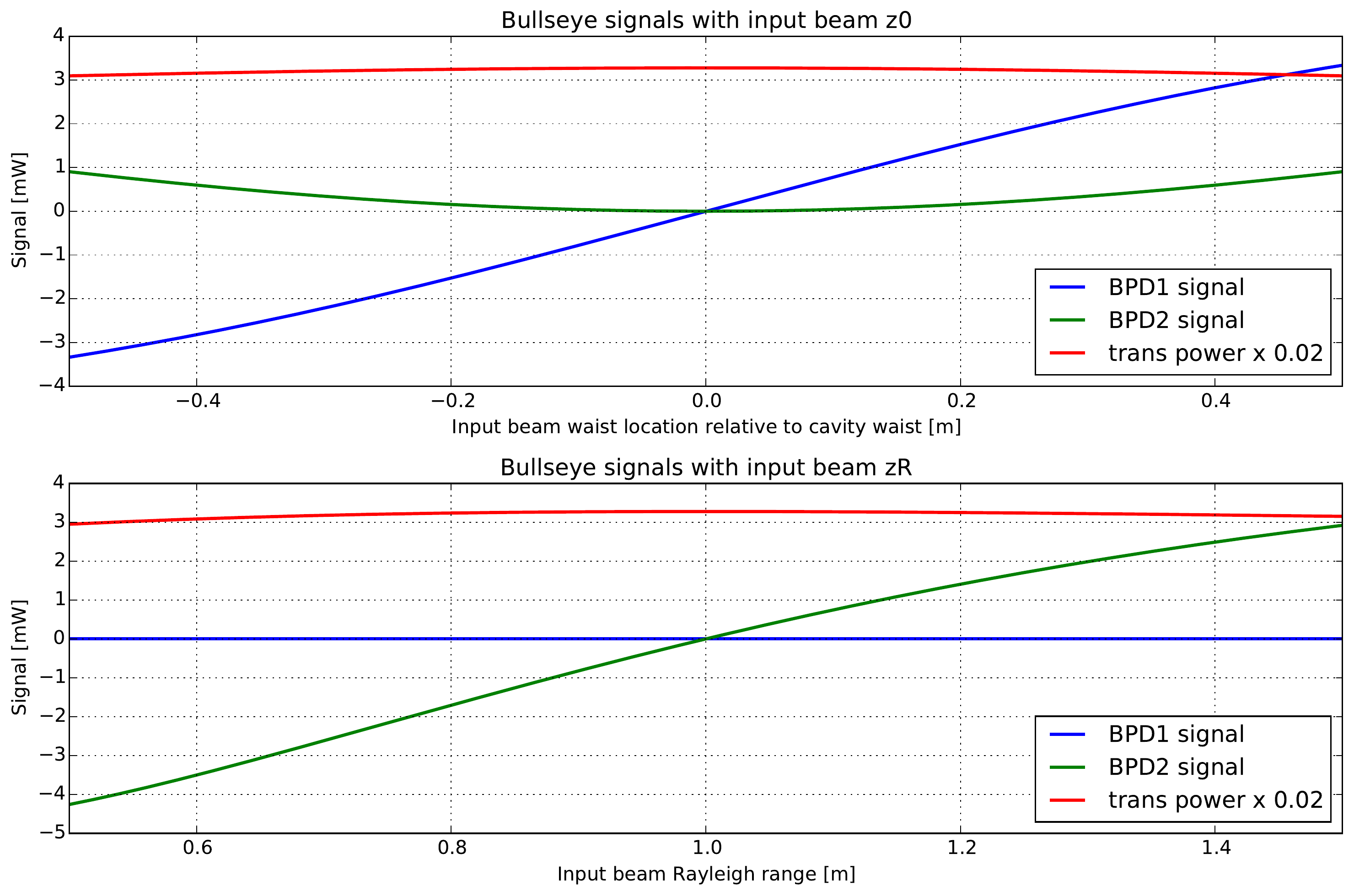}
\caption{Simulated bullseye detector signals as a function of cavity input beam parameters, for a fixed 
cavity eigenmode. BPD1 is primarily sensitive to the input beam waist location mismatch, whereas BPD2 is primarily 
sensitive to input beam waist size (or Rayleigh range) mismatch.}
\label{fig:BPDplots}
\end{figure}
\end{center}

The output signals from the two bullseye detectors are plotted in Fig.~\ref{fig:BPDplots} as a function of the input beam waist location 
relative to the cavity waist (upper plot) and the input beam Rayleigh range (lower plot). The cavity transmitted light power 
is also shown in the plots. Here we see that BPD1 is primarily sensitive to the input beam waist location mismatch, whereas BPD2 is primarily 
sensitive to input beam waist size (or Rayleigh range) mismatch. We also see that both bullseye detector signals are 
linear in the region close to optimal mode matching, confirming that these would make appropriate error signals 
for use in a closed-loop feedback control system for mode matching the input beam to the cavity mode. 

\section{Code to compute coefficients}
The computation of beat coefficients for split and bullseye detectors
has been added as a feature to our Python-based \textsc{PyKat}
package~\cite{pykat}. To generate the default coefficients for
\textsc{Finesse} to be stored in the kat.ini file, install \textsc{Pykat}
and run:
\begin{verbatim}
import pykat.optics.pdtype as pdtype
pdtype.finesse_bullseye_photodiode(6)
\end{verbatim}

\newpage
\bibliographystyle{unsrt}
\bibliography{BPDbib}


\end{document}